\documentclass[10pt]{article}
\usepackage{amsmath,amssymb,amsfonts,amsthm,graphicx}
\title{Nonclassical properties of coherent states and excited coherent states for continuous spectra}
\author{G.R. Honarasa$^{1,2}$,
M.K. Tavassoly$^1$, M. Hatami$^1$ and R. Roknizadeh$^3$
\\
\footnotesize{$^1$ Atomic and Molecular Group, Faculty  of Physics, Yazd University, Yazd, Iran }
\\
\footnotesize{ $^2$ Physics Group, Faculty of Science, Shiraz University of Technology, Shiraz, Iran} \\
\footnotesize{ $^3$ Quantum Optics Group, Department of Physics, University of Isfahan, Isfahan, Iran}
\\ \footnotesize{e-mail: mktavassoly@yazduni.ac.ir  } }

\begin{document}

\maketitle \thispagestyle{empty}

 \begin{abstract}
   Based on the definition of coherent states for continuous spectra and analogous to photon added coherent
  states for discrete spectra, we introduce the excited coherent states for continuous spectra. It is shown
  that, the main axioms of Gazeau-Klauder coherent states will be satisfied, properly. Nonclassical properties
  and quantum statistics of coherent states, as well as the introduced excited coherent states are discussed.
  In particular, through the study of quadrature squeezing and amplitude squared squeezing, it will be observed that both
  classes of the above states can be classified in the intelligent states category.

 \end{abstract}

 {\bf keyword:}
   Coherent states, Excited coherent states, Continuous spectra, Mandel parameter, Quadrature squeezing

{\it PACS:} 03.65.-w, 42.50.-p

\newcommand{\I}{\mathbb{I}}
\newcommand{\norm}[1]{\left\Vert#1\right\Vert}
\newcommand{\abs}[1]{\left\vert#1\right\vert}
\newcommand{\set}[1]{\left\{#1\right\}}
\newcommand{\R}{\mathbb R}
\newcommand{\C}{\mathbb C}
\newcommand{\DD}{\mathbb D}
\newcommand{\eps}{\varepsilon}
\newcommand{\To}{\longrightarrow}
\newcommand{\BX}{\mathbf{B}(X)}
\newcommand{\HH}{\mathfrak{H}}
\newcommand{\A}{\mathcal{A}}
\newcommand{\N}{\mathcal{N}}
\newcommand{\B}{\mathcal{B}}
\newcommand{\RR}{\mathcal{R}}
\newcommand{\HD}{\hat{\mathcal{H}}}

  \section{Introduction}\label{sec-intro}
Coherent states play an important role in various fields of physics, specially in quantum technologies and quantum
 optics \cite{klauder1,ali,deuar,chong}. 
 %
  Gazeau and Klauder introduced coherent states for systems with either
  discrete, continuous or both discrete and continuous spectra \cite{gazeau}. Later, ladder operator structure
  of discrete type of the latter states has been established in \cite{tavassoly}. Recently, a theoretical scheme for generation of Gazeau-Klauder coherent states via intensity-dependent degenerate Raman interaction and also Gazeau-Klauder squeezed states associated with solvable quantum systems with discrete spectra have been introduced and discussed by one of us \cite{yadollahi,tavassoly1}.\\
Recently, Ben Geloun and Klauder introduced two kinds of ladder operators for quantum systems with continuous spectra
\cite{geloun}. There, two distinct classes of coherent states were introduced (but, only one of them, which has been constructed based on the eigenstate definition of annihilation operator, is appropriate for the goal of the present work). Then, the authors established the axioms of the Gazeau-Klauder coherent states on their introduced states. Moreover, the main motivation for generalizing the coherent states lies in finding the nonclassical properties (states with no classical analogue). This is due to the fact that, nonclassical states have received considerable attention in the fields of quantum optics, quantum cryptography and quantum communication, etc \cite{klauder1,ali,kampe,bennett,schumacher}. Altogether, neither of the authors previously dealt with coherent states for continuous spectra \cite{gazeau, geloun} pay attention to this important subject, which is of enough interest for quantum optics specialists.\\ 
It is worth to notice that the notions of "coherent states" and "nonclassicality of states" are not restricted to radiation field. As some cases we may refer to fermion coherent states \cite{klauder1}, coherent spin states \cite{rad}, coherent states for potentials other than harmonic oscillator \cite{Nieto1,Nieto2} and nonlinear coherent states of the center of mass motion of a trapped ion \cite{filho}. Atomic spin-squeezed-states and the associated spin squeezing parameter \cite{Civitarese} is a well-known example that squeezing as a nonclassicality sign does not necessarily related to the radiation field. Along these, we have generalized some of the nonclassicality criteria to the coherent states for continuous spectra. \\
 In the present paper, we will briefly review the definition of one kind of ladder operators and the associated coherent states
 for continuous spectra proposed by Ben Geloun and Klauder, which satisfy the eigenvalue problem, including a few additional
 remarkable points in section 2. Then, analogous to photon added coherent states were introduced by Agarwal and Tara
 \cite{agarwal}, the excited coherent states corresponding to continuous spectra will be introduced in section 3. Also,
 the axioms of Gazeau-Klauder coherent states will be examined for the introduced states. Next, some nonclassicality
 features of the coherent and excited coherent states for continuous spectra are investigated, respectively in sections
 4 and 5. Finally, we conclude the paper in section 6.
 \section{Annihilation and creation operators and the associated coherent states for continuous spectra}
Let us suppose that the dynamics of a system is described by the Hamiltonian $H$, which admits a non-degenerate continuous
spectrum. The corresponding eigenstates are denoted by $|E\rangle$, where they satisfy the following eigenvalue equation
and orthogonality relation, respectively as
\begin{equation}\label{HE}
   H|E\rangle=\omega E|E\rangle,\;\;\;\;\langle E|E^\prime\rangle=\delta (E-E^\prime),
\end{equation}
in which $E>0$ and $\hbar=1$ is assumed. Along the study of coherent states for such systems, Ben Geloun and Klauder
introduced the following annihilation operator on the continuous basis \cite{geloun}
\begin{equation}\label{aepsilon}
   a_\epsilon=\int_0^\infty C(E,\epsilon)|E-\epsilon\rangle \langle E| dE,\qquad C(E,\epsilon)=e^{\alpha(\epsilon E-\frac{1}{2}\epsilon^2)},
\end{equation}
where $\epsilon>0$ and $\alpha$ are real parameters. The operator $a_\epsilon$ which decreases $E$ by an amount of energy $\epsilon$ may be called an "energy depleting operator". They then introduced the coherent states
 $|s,\gamma \rangle_\epsilon$ satisfying the eigenvalue problem
\begin{equation}\label{asgamma}
   a_\epsilon|s,\gamma\rangle_\epsilon=(se^{-i\gamma})^\epsilon|s,\gamma\rangle_\epsilon,
\end{equation}
where $s\in [0,\infty)$ and $\gamma \in (-\infty,\infty)$. The
following expression for the associated coherent states on the
continuous basis has been obtained
\begin{equation}\label{sgamma}
   |s,\gamma\rangle_\epsilon=\N_\epsilon(s)\int_0^\infty \frac{s^E}{e^{\frac{1}{2}\alpha E^2}}e^{-i\gamma E}|E\rangle dE,
\end{equation}
where
\begin{equation}\label{normalization factor}
   \N_\epsilon(s)=\left[\int_0^\infty \frac{s^{2E}}{e^{\alpha E^2}} dE\right]^{-\frac{1}{2}},
\end{equation}
is a normalization factor. The states (\ref{sgamma}) are the same as those determined by Gazeau-Klauder
 for a given function $f(E)=e^{\frac{1}{2}\alpha E^2}$ of Ref. \cite{gazeau}. Creation operator or "energy supplying operator" can be simply obtained by Hermitian conjugation of (\ref{aepsilon}) with the result
\begin{equation}\label{adaggerepsilon}
   a_\epsilon^\dagger=\int_0^\infty C(E,\epsilon)|E\rangle \langle E-\epsilon| dE,
\end{equation}
where $C(E,\epsilon)$ is given by (\ref{aepsilon}). Both annihilation and creation operators respectively introduced in (\ref{aepsilon}) and (\ref{adaggerepsilon}) act to either decrease or increase the energy scale by a uniform amount $\epsilon$ (i.e., an amount independent of the energy $E$ itself). Although, these operators have energy dependent (the weighting factor $C(E,\epsilon)$), but this effects the "efficiency" with which the subtracted or added energy is received by this particular "energy transforming device". \\
The ladder operators $a_\epsilon$ and $a_\epsilon^\dagger$
 obey the following commutation relation
 \begin{eqnarray}\label{a,adagger}
   [a_\epsilon,a_\epsilon^\dagger]&=&\int_0^\infty \left(C^2(E+\epsilon,\epsilon)-C^2(E,\epsilon)\right)|E\rangle \langle E|dE \nonumber \\
 &=&2\sinh{(\alpha \epsilon^2)}\int_0^\infty e^{2\alpha \epsilon E}|E\rangle \langle E|dE.
\end{eqnarray}
 While the rising and lowering operators on the continuous basis are defined by Ben Geloun and Klauder in \cite{geloun},
  we can now introduce a number-like operator on the same basis with the help of (\ref{aepsilon}) and (\ref{adaggerepsilon}) as follows
 \begin{equation}\label{Nepsilon}
   N_\epsilon=a_\epsilon^\dagger a_\epsilon=\int_0^\infty e^{2\alpha(\epsilon E-\frac{1}{2}\epsilon^2)} |E\rangle \langle E| dE.
\end{equation}
The physical meaning of this operator is not the same as number operator in the discrete case, because when the usual annihilation and creation operators act on discrete states, one speaks of decreasing or increasing "particles", i.e., photons, e.g., from the fact that $| n \rangle \rightarrow | n-1 \rangle$ and the converse, but in the continuous case the "energy" is decreased or increased by a uniform amount.\\
The following eigenvalue equation holds:
\begin{equation}\label{NE}
 N_\epsilon |E\rangle=e^{-\alpha \epsilon^2}e^{2\alpha \epsilon E} |E \rangle.
 \end{equation}
The number-like operator satisfy the following commutators with $a_\epsilon$ and $a_\epsilon^\dagger$, respectively
\begin{eqnarray}\label{N,a}
   [N_\epsilon,a_\epsilon]&=&\int_0^\infty C(E,\epsilon) \left(C^2(E-\epsilon,\epsilon)-C^2(E,\epsilon)
   \right)|E-\epsilon\rangle \langle E|dE \nonumber \\
 &=&-\int_0^\infty e^{3\alpha \epsilon E-\frac{7}{2}\alpha \epsilon^2}(e^{2\alpha \epsilon^2}-1)|E-\epsilon\rangle \langle E|dE,
\end{eqnarray}
\begin{eqnarray}\label{N,adagger}
   [N_\epsilon,a_\epsilon^\dagger]&=&\int_0^\infty C(E,\epsilon) \left(C^2(E,\epsilon)-C^2(E-\epsilon,
   \epsilon)\right)|E\rangle \langle E-\epsilon|dE \nonumber \\
 &=&\int_0^\infty e^{3\alpha \epsilon E-\frac{7}{2}\alpha \epsilon^2}(e^{2\alpha \epsilon^2}-1)|E\rangle \langle E-\epsilon|dE.
\end{eqnarray}
Therefore, the operators $a_\epsilon$, $a_\epsilon^\dagger$ and $N_\epsilon$ do not obey the standard
 Weyl-Heisenberg algebra.  Ben Geloun and Klauder obtained the following expression at limit
\begin{equation}\label{I}
   \lim_{\alpha\rightarrow0}\frac{[a_\epsilon,a_\epsilon^\dagger]}{2\alpha \epsilon^2}=\int_0^\infty |E\rangle \langle E| dE=\I,
 \end{equation}
 where $\I$ is the unity operator in the quantum Hilbert space. One may continue with calculating the
 commutators between  $N_\epsilon$ with $a_\epsilon$ and
 $a_\epsilon^\dagger$, respectively as
 \begin{equation}\label{lima}
   \lim_{\alpha\rightarrow0}\frac{[N_\epsilon,a_\epsilon]}{2\alpha \epsilon^2}=-\int_0^\infty
   |E\rangle \langle E-\epsilon| dE=-\lim_{\alpha\rightarrow0} a_\epsilon,
 \end{equation}
 \begin{equation}\label{limadagger}
   \lim_{\alpha\rightarrow0}\frac{[N_\epsilon,a_\epsilon^\dagger]}{2\alpha \epsilon^2}=
   \int_0^\infty |E-\epsilon\rangle \langle E| dE=\lim_{\alpha\rightarrow0} a_\epsilon^\dagger.
 \end{equation}
  So, we can conclude that the set $\{a_\epsilon, a_\epsilon^\dagger, N_\epsilon, \I\}$ constitute the
  generators of a deformed version of Heisenberg algebra.
 \section{Excited coherent states for continuous spectra}
 Photon added coherent states were introduced by Agarwal and Tara \cite{agarwal} by the iterated
 actions ($m$ times) of $a^\dagger$ on the canonical coherent states, i.e.,
 \begin{equation}\label{photon add}
 |\alpha,m\rangle=\frac{{a^\dagger}^m|\alpha\rangle}{\sqrt{\langle \alpha|a^m {a^\dagger}^m|\alpha \rangle}},
 \end{equation}
  where $a$ and $a^\dagger$ are bosonic annihilation and creation operators, respectively and $|\alpha
  \rangle$ is the canonical coherent state. Later, photon added coherent states for solvable quantum systems
  with discrete spectra were introduced in \cite{daoud}. Recently, photon added coherent states were generated
  experimentally by Zavatta et al \cite{zavata}. Analogously, we are now interested in the introduction of
  excited coherent states for continuous spectra as follows:
 \begin{equation}\label{energy add}
 |s,\gamma,m\rangle_\epsilon=\frac{{a_\epsilon^\dagger}^m|s,\gamma\rangle_\epsilon}{\sqrt{_\epsilon \langle
  s,\gamma|a_\epsilon^m {a_\epsilon^\dagger}^m|s,\gamma \rangle_\epsilon}},
 \end{equation}
 where $m$ is a non-negative integer. With the help of (\ref{sgamma}) and (\ref{adaggerepsilon}), these states get the following form
 \begin{equation}\label{energy added0}
 |s,\gamma,m\rangle_\epsilon=\N_{\epsilon,m}(s) \int_0^\infty \left(\prod_{j=0}^{m-1}e^{\alpha[\epsilon
 E+(m-j-\frac{1}{2})\epsilon^2]}\right)\frac{s^E e^{-i\gamma E}}{e^{\frac{1}{2}\alpha E^2}}|E+m\epsilon \rangle
 dE,
 \end{equation}
 in which we have set $\N_{\epsilon,m}(s)=({_\epsilon \langle
  s,\gamma|a_\epsilon^m {a_\epsilon^\dagger}^m|s,\gamma \rangle_\epsilon})^{-1/2}$. Due to the relation $\prod_{j=0}^{m-1}e^{\alpha[\epsilon E+(m-j-\frac{1}{2})\epsilon^2]}=e^{\frac{1}{2}m\alpha
 \epsilon(2E+m\epsilon)}$, the relation (\ref{energy added0}) simplifies to
 \begin{equation}\label{energy added}
 |s,\gamma,m\rangle_\epsilon=\N_{\epsilon,m}(s) \int_0^\infty e^{\frac{1}{2}m\alpha \epsilon(2E+m\epsilon)}
 \frac{s^E e^{-i\gamma E}}{e^{\frac{1}{2}\alpha E^2}}|E+m\epsilon \rangle dE.
 \end{equation}
 The normalization factor, $\N_{\epsilon,m}(s)$, can be obtained by requiring $_\epsilon \langle s,\gamma,m|s,\gamma,m\rangle_\epsilon=1$ and $\alpha>0$ as
 \begin{eqnarray}\label{Ns,m}
 \N_{\epsilon,m}(s)&=&\left[\int_0^\infty e^{m\alpha \epsilon(2E+m\epsilon)}\frac{s^{2E}}{e^{\alpha E^2}}dE\right]
 ^{-\frac{1}{2}} \nonumber \\
 &=&(\frac{4\alpha}{\pi})^{\frac{1}{4}}e^{-\frac{(\ln s)^2}{2\alpha}} e^{-m^2\alpha \epsilon^2} s^{-m\epsilon}
  \left[1+\mathrm{erf}\left(\frac{m\alpha \epsilon+\ln s}{\sqrt{\alpha}}\right)\right]^{-\frac{1}{2}},
 \end{eqnarray}
 where erf(.) represents the Gaussian error function. Setting $m=0$ in (\ref{energy added}) and (\ref{Ns,m})
  the state $|s,\gamma,m\rangle_\epsilon$ reduces to coherent state for continuous spectra in (\ref{sgamma})
  and (\ref{normalization factor}).\\
 At this point we want to check Gazeau-Klauder axioms for the introduced excited coherent states corresponding
 to continuous spectra. The main axioms are as follows.\\
 (a) {\it Continuity of labelling:} it is clearly satisfied.\\
 (b) {\it Temporal stability:} using (\ref{energy added}) and the relevant Hamiltonian (\ref{HE}) readily gives
 \begin{eqnarray}\label{temporal}
 e^{-iHt}|s,\gamma,m\rangle_\epsilon &=&e^{-im\epsilon \omega t}\N_{\epsilon,m}(s) \int_0^\infty e^{\frac{1}{2}m\alpha
 \epsilon(2E+m\epsilon)} \frac{s^E e^{-i(\gamma+\omega t) E}}{e^{\frac{1}{2}\alpha E^2}}|E+m\epsilon \rangle dE  \nonumber  \\
&=&e^{-im\epsilon \omega t} |s,\gamma+\omega t,m\rangle_\epsilon.
 \end{eqnarray}
 (c) {\it Resolution of the identity:}
 \begin{equation}\label{res1}
 \int_{-\infty} ^\infty \frac{d\gamma}{2\pi} \int_0^\infty ds \sigma_m(s) |s,\gamma,m\rangle_\epsilon\; _\epsilon\langle s,\gamma,m|=\I_m,
 \end{equation}
 where we have defined the unity operator, $\I_m$, as
 \begin{equation}\label{Im}
 \I_m=\int_0^\infty |E+m\epsilon\rangle \langle E+m\epsilon|dE,
 \end{equation}
 analogously to the unity operator has been introduced for photon added coherent states of discrete
 spectra \cite{penson}.
 With the help of (\ref{energy added}) and performing the integral on $\gamma$, using $\int_{-\infty}
  ^\infty e^{i\gamma(E-E^\prime)}d\gamma=2\pi \delta(E-E^\prime)$ in the left hands side of (\ref{res1}),
   finally leads one to the Stieljes moment problem
 \begin{equation}\label{res2}
 \int_0^\infty ds\; h_m(s) s^{2E}=e^{\alpha E^2} e^{-m\alpha \epsilon(2E+m\epsilon)}=e^{\alpha (E-m\epsilon)^2} e^{-2m^2\alpha \epsilon^2},
 \end{equation}
 where $h_m(s)\equiv \sigma_m(s)[\N_{\epsilon,m}(s)]^2$. Considering new variables $u=\ln s$ and
 $E^\prime=E-m\epsilon$ this integral equation converted to
 \begin{equation}\label{res3}
 \int_{-\infty}^\infty du\; \tilde{h}_m(u) e^{2E^\prime u}=e^{\alpha {E^\prime}^2}e^{-2m^2\alpha \epsilon^2} ,
 \end{equation}
with the solution $\tilde{h}_m(u)=e^{-2m^2\alpha \epsilon^2}
e^{-\frac{u^2}{\alpha}}/\sqrt{\alpha \pi}$. So,  we have
 \begin{equation}\label{res4}
 \sigma_m(s)=\frac{1}{s^{2m\epsilon+1}\sqrt{\alpha \pi}} e^{-\frac{(\ln s)^2}{\alpha}}e^{-2m^2\alpha \epsilon^2} [\N_{\epsilon,m}(s)]^{-2}.
 \end{equation}
 Using (\ref{Ns,m}), $\sigma_m(s)$ may be finally written in the following closed form
 \begin{equation}\label{res5}
 \sigma_m(s)=\frac{1}{2s\alpha} \left[1+\mathrm{erf}\left(\frac{m\alpha \epsilon+\ln s}{\sqrt{\alpha}}\right)\right].
 \end{equation}
 The excited coherent state will be coherent in its exact meaning, if $\sigma_m(s)\geq 0$ exists for
 all ranges of $s$. To test this requirement, in figure 1 we have displayed the function $\sigma_m(s)$ for different values of $m$.
 It is seen that $\sigma_m(s)$ is positive for all values of $s$ and $\lim_{s\rightarrow \infty}\sigma_m(s)=0$.\\
 (d) {\it Action identity:} the action identity can be deduced from the mean value of the Hamiltonian
  over the excited states, which yields
 \begin{eqnarray}\label{action1}
 _\epsilon\langle s,\gamma,m|H|s, \gamma, m\rangle_\epsilon= \omega [\N_{\epsilon,m}(s)]^{2} \int_0^\infty
  e^{m\alpha \epsilon(2E+m\epsilon)} \frac{s^{2E}}{e^{\alpha E^2}} E\;dE \equiv\omega J(s),
 \end{eqnarray}
 where the new action variable is assumed to be invertible versus $s$. As argued in \cite{gazeau}, if the
 function $J(s)$ is invertible such that $s(J)$ can be determined, the excited coherent states (\ref{energy added})
  satisfy the action identity axiom, appropriately.\\
So, we have established that the excited coherent states for continuous spectra (\ref{energy added})
 maintain in the family of Gazeau-Klauder type.

 \section{Nonclassical properties of "coherent states" for continuous spectra}
  Before investigating the nonclassicality features of our introduced excited states,
  in this section, some of the nonclassical properties of "coherent states" for continuous spectra (\ref{sgamma})
  will be studied. For this purpose, we will discuss Mandel parameter, second-order correlation function,
  quadrature squeezing and amplitude-squared squeezing.
 \subsection{Mandel parameter}
Several parameters were introduced to characterize the statistical properties. The most
popular one among them is Mandel parameter $Q$, which was frequently used to measure the deviation from Poissonian
distribution  \cite{mandel}. Calculating the mean value of $N_\epsilon$ with respect to the states (\ref{sgamma}) yields
\begin{equation}\label{Nepsilonavr}
 \langle N_\epsilon \rangle=\,  _\epsilon \langle s,\gamma |N_\epsilon |s,\gamma \rangle_\epsilon =s^{2\epsilon}\;.
\end{equation}
Also, noticing that
\begin{equation}\label{N2epsilon}
   N_\epsilon^2=\int_0^\infty \int_0^\infty e^{4\alpha(\epsilon E-\frac{1}{2}\epsilon^2)} |E\rangle \langle E| dE,
\end{equation}
we get
\begin{equation}\label{N2epsilon2}
\langle N_\epsilon^2 \rangle =\,_\epsilon \langle s,\gamma |N_\epsilon^2 |s,\gamma \rangle_\epsilon=e^{2\alpha \epsilon^2}s^{4\epsilon}.
\end{equation}
Therefore, extending the Mandel parameter definition to continuous
spectra seems to be possible \cite{mandel}, i.e.,
\begin{equation}\label{Q}
Q_\epsilon=\frac{\langle N_\epsilon^2\rangle-\langle N_\epsilon \rangle^2}{\langle N_\epsilon\rangle}-1.
\end{equation}
The $Q_\epsilon$ quantity determines the quantum statistics of the
radiation field states for continuous spectra, i.e., it is
super-Poissonian (if $Q_\epsilon > 0$), sub-Poissonian (if
$Q_\epsilon < 0$) or Poissonian (if $Q_\epsilon = 0$). Using
(\ref{Nepsilonavr}) and (\ref{N2epsilon2}) in (\ref{Q}), the
Mandel parameter for $|s,\gamma\rangle_\epsilon$ states will be
obtained in closed form as follows
\begin{equation}\label{Q1}
Q_\epsilon=s^{2\epsilon}(e^{2\alpha \epsilon^2}-1)-1.
\end{equation}
It is noticeable that all of the quantities $\langle N_\epsilon
\rangle$, $\langle N_\epsilon^2 \rangle$ and at last $Q_\epsilon$,
which obtained in closed form, are independent of $\gamma$. In
figure 2, we have plotted Mandel parameter of coherent states for
continuous spectra versus $s$ for different values of $\epsilon$,
 keeping $\alpha$ fixed at $10$. The figure shows that the coherent states for continuous spectra obey sub-Poissonian
  statistics ($Q_\epsilon<0$) for some intervals of $s$. From the figure, it is seen that as $\epsilon$ decreases the
  interval of $s$, for which sub-Poissonian statistics reveals will be increased. For small values of $\epsilon$,
  Mandel parameter is almost equal to $-1$. This is the value of Mandel parameter corresponding to the number states,
  as the most nonclassical states. Noticing that in the presented approach $\epsilon>0$ is
   the continuity parameter, one can be sure that for really continuous states the
   sub-Poissonian behavior of the states may not become weaker than what we have shown in figures.
   In figure 3 we have displayed
   Mandel parameter of coherent states for continuous spectra versus $s$ for different
   values of $\alpha$, keeping $\epsilon$ fixed at $0.07$. From the figure, we find that
    smaller $\alpha$ shows more nonclassical aspects.
 \subsection{Second-order correlation function}
By developing the second-order correlation function definition \cite{scully} to quantum systems with continuous spectra, we define:
\begin{equation}\label{g20}
g_\epsilon^2(0)=\frac{\langle {a_\epsilon^\dagger}^2 a_\epsilon^2\rangle}{\langle a_\epsilon^\dagger a_\epsilon \rangle^2}
\end{equation}
If $g_\epsilon^2(0)<1$ ($g_\epsilon^2(0)\geq1$) the state exhibits nonclassical (classical) behavior. With respect
 to the coherent states associated with continuous spectra (\ref{sgamma}), the following mean value analytically obtained
 \begin{equation}\label{adager2a2}
\langle {a_\epsilon^\dagger}^2 a_\epsilon^2 \rangle =\,_\epsilon \langle s,\gamma |{a_\epsilon^\dagger}^2
a_\epsilon^2 |s,\gamma \rangle_\epsilon=s^{4\epsilon}.
\end{equation}
Using (\ref{Nepsilonavr}) and (\ref{adager2a2}), second-order correlation function for coherent states of
continuous spectra will be exactly $1$, just as in the canonical coherent states. Therefore, coherent states for continuous spectra do not show this nonclassical behavior.

 \subsection{Quadrature squeezing}
 As another criteria for nonclassicality of states, we study quadrature squeezing  \cite{walls} of
  the coherent states in (\ref{sgamma}). For this purpose, let us consider the following Hermitian quadrature operators
 \begin{equation}\label{x}
 X_1 =\frac{a_\epsilon+a_\epsilon^\dagger}{2},\;\;\;\;Y_1 =\frac{a_\epsilon-a_\epsilon^\dagger}{2i}.
\end{equation}
 The expectation values of $X_1$ and $X_1^2$ with respect to the states (\ref{sgamma}) are respectively given analytically by
 \begin{equation}\label{x1}
\langle X_1 \rangle =\,_\epsilon \langle s,\gamma |X_1 |s,\gamma \rangle_\epsilon=s^\epsilon \cos{(\gamma \epsilon)}
\end{equation}
and
\begin{equation}\label{x12}
\langle X_1^2 \rangle =\,_\epsilon \langle s,\gamma |X_1^2 |s,\gamma \rangle_\epsilon=\frac{1}{2}s^{2\epsilon}
\cos{(2\gamma \epsilon)}+\frac{1}{4}s^{2\epsilon} (e^{2\alpha \epsilon^2}+1).
\end{equation}
Thus, the dispersion of $X_1$ may be written in the closed form as
\begin{equation}\label{deltax1}
(\Delta X_1)^2=\langle X_1^2 \rangle-\langle X_1 \rangle^2 =\frac{1}{4}s^{2\epsilon} (e^{2\alpha \epsilon^2}-1).
\end{equation}
The same result is obtained for the dispersion of operator $Y_1$,
\begin{equation}\label{deltay1}
(\Delta Y_1)^2=\langle Y_1^2 \rangle-\langle Y_1 \rangle^2 =\frac{1}{4}s^{2\epsilon} (e^{2\alpha \epsilon^2}-1).
\end{equation}
Also, due to the relation
\begin{equation}\label{x1y10}
[X_1,Y_1]=\frac{i}{2}[a_\epsilon,a_\epsilon^\dagger]=i\sinh (\alpha \epsilon^2)\int_0^\infty e^{2 \alpha \epsilon E}|E\rangle \langle E|dE,
\end{equation}
one readily may arrive at
\begin{equation}\label{x1y1}
\langle [X_1,Y_1] \rangle =\,_\epsilon \langle s,\gamma |[X_1,Y_1] |s,\gamma \rangle_\epsilon=\frac{i}{2}s^{2\epsilon}(e^{2\alpha \epsilon^2}-1),
\end{equation}
Therefore, it is clear that $X_1$ and $Y_1$ minimize the Heisenberg uncertainty relation. Indeed,
we have equal fluctuations in $X_1$ and $Y_1$, and
\begin{equation}\label{x1y1uncern}
(\Delta X_1)^2(\Delta Y_1)^2 =\frac{1}{4} \left|\langle [X_1,Y_1] \rangle \right|^2.
\end{equation}
Clearly, no squeezing may be expected. It is worth to mention
that, in a sense, with respect to the quadratures in (\ref{x}),
the states defined in (\ref{sgamma}) are of "intelligent type"
states \cite{tavassoly2,vogel}.
 \subsection{Amplitude-squared squeezing}
 In order to examine whether the coherent states (\ref{sgamma}) exhibit higher-order squeezing,
  particularly amplitude-squared squeezing, or not, we introduce the following Hermitian operators
\begin{equation}\label{y}
 X_2 =\frac{a_\epsilon^2+{a_\epsilon^\dagger}^2}{2},\;\;\;\;Y_2 =\frac{a_\epsilon^2-{a_\epsilon^\dagger}^2}{2i}.
\end{equation}
In this case, the dispersions are straightforwardly obtained in closed form as
\begin{equation}\label{deltax2y2}
(\Delta X_2)^2=(\Delta Y_2)^2=\frac{1}{4}s^{4\epsilon}(e^{8\alpha \epsilon^2}-1).
\end{equation}
In general, for the commutator of $X_2$ and $Y_2$ one gets
\begin{equation}\label{x2y20}
[X_2,Y_2]=\frac{i}{2}[a_\epsilon^2,{a_\epsilon^\dagger}^2]=i\sinh (4\alpha \epsilon^2)\int_0^\infty e^{4 \alpha \epsilon E}|E\rangle \langle E|dE,
\end{equation}
from which we obtain
\begin{equation}\label{x2y2}
\langle [X_2,Y_2] \rangle =\,_\epsilon \langle s,\gamma |[X_2,Y_2] |s,\gamma \rangle_\epsilon=\frac{i}{2}s^{4\epsilon}(e^{8\alpha \epsilon^2}-1).
\end{equation}
Hence one has
\begin{equation}\label{x2y2uncern}
(\Delta X_2)^2(\Delta Y_2)^2 =\frac{1}{4} \left|\langle [X_2,Y_2] \rangle \right|^2,
\end{equation}
and therefore, no squeezing occurs, even in amplitude-squared
operators and they are also minimum uncertainty or intelligent
states. It is noticeable that, neither of the expectation values
calculated in the first and second order squeezing depend on
$\gamma$.
 \section{Nonclassical properties of "excited coherent states" for continuous spectra}
 In this section, the nonclassicality of the introduced "excited coherent states" for continuous spectra (\ref{energy added})
  will be investigated, using the same criteria have been used in the previous section.
 \subsection{Mandel parameter}
 By the definition of number-like operator (\ref{Nepsilon}), the following mean values for excited coherent states are analytically obtained:
 \begin{equation}\label{Nenergyadded}
 \langle N_{\epsilon, m} \rangle=\, _\epsilon \langle s,\gamma,m |N_\epsilon |s,\gamma,m \rangle_\epsilon  =s^{2\epsilon}e^{4m\alpha  \epsilon^2},
 \end{equation}
 and
 \begin{equation}\label{N2energy added}
\langle N_{\epsilon, m}^2 \rangle =\,_\epsilon \langle s,\gamma,m
|N_\epsilon^2 |s,\gamma,m \rangle_\epsilon=e^{2\alpha
\epsilon^2(1+4m)}s^{4\epsilon}.
\end{equation}
Finally, Mandel parameter gets the following closed form
\begin{equation}\label{Qenergyadded}
Q_{\epsilon, m}=s^{2\epsilon}e^{4m\alpha \epsilon^2}(e^{2\alpha
\epsilon^2}-1)-1,
\end{equation}
where the subscript $m$ indicates to the order of excited coherent
states. It is seen that $Q_{\epsilon,m}$ is also
$\gamma$-independent. Notice that setting $m=0$ in
(\ref{Qenergyadded}) recovers (\ref{Q1}). Figure 4 shows the plots
of Mandel parameter of excited coherent states for continuous
spectra versus $s$
 for different values of $\epsilon$, fixing $\alpha=3$ and $m=2$. Also, we have plotted Mandel parameter
 of excited coherent states versus $s$ for different values of $\alpha$, while keeping $\epsilon$ fixed at $0.07$
 in figure 5. The figures show sub-Poissonian statistics for some intervals of $s$.
 Our further calculations show that these intervals will be wider as $\epsilon$ or
  $\alpha$ decrease. In figure 6, Mandel parameter of excited coherent states for continuous
   spectra has been plotted versus $s$ for various values of $m$. From the figure we find
   that, unlike the photon added coherent states in \cite{agarwal}, nonclassicality depth
    and range will be decreased with increasing $m$.
 \subsection{Second-order correlation function}
 Now, we can also study second-order correlation function for
 excited coherent states associated with continuous spectra,
 (\ref{energy added}). For these states one analytically obtains
 \begin{equation}\label{adager2a22}
  \langle {a_\epsilon^\dagger}^2 a_\epsilon^2 \rangle =\,_\epsilon
  \langle s,\gamma,m |{a_\epsilon^\dagger}^2 a_\epsilon^2
  |s,\gamma,m \rangle_\epsilon=s^{4\epsilon}e^{8m\alpha \epsilon^2}.
\end{equation}
 With the help of (\ref{Nenergyadded}) for $\langle a_\epsilon^\dagger a_\epsilon \rangle$, clearly one has,
 just as the coherent states for continuous spectra and canonical coherent states, $g_{\epsilon,m}^2(0)=1$.
 This implies that, these states do not exhibit nonclassical behavior (independent on the values of $m$ and $\epsilon$),
 while they possess sub-Poissonian statistics.
 \subsection{Quadrature squeezing}
 With the help of the definitions of Hermitian operators $X_1$ and $Y_1$ introduced in (\ref{x}), the dispersions
  with respect to the excited coherent states in (\ref{energy added}) are given in closed form by
 \begin{equation}\label{deltax2y2energyadded}
(\Delta X_1)_m^2=(\Delta Y_1)_m^2=\frac{1}{4}s^{2\epsilon}e^{4m\alpha \epsilon^2}(e^{2\alpha \epsilon^2}-1).
\end{equation}
It follows also that
\begin{equation}\label{x1y1energy added}
\langle [X_1,Y_1] \rangle_m =\, _\epsilon \langle s,\gamma,m |[X_1,Y_1] |s,\gamma,m \rangle_\epsilon=\frac{i}{2}
s^{2\epsilon}e^{4m\alpha \epsilon^2}(e^{2\alpha \epsilon^2}-1),
\end{equation}
 and again one has
 \begin{equation}\label{nosqueeze0}
(\Delta X_1)_m^2=(\Delta Y_1)_m^2=\frac{1}{2}|\langle [X_1,Y_1] \rangle_m|.
\end{equation}
Therefore, no quadrature squeezing occurs for our introduced excited states and they saturate the Heisenberg
uncertainty relation. These states may also be called as intelligent states, regarding $X_1$ and $Y_1$ quadratures.
Again, putting $m=0$ in (\ref{deltax2y2energyadded}) and (\ref{x1y1energy added}) reduce to (\ref{deltax1}) and (\ref{x1y1}), respectively.
 \subsection{Amplitude-squared squeezing}
 By using (\ref{y}) and (\ref{energy added}), the following expressions for dispersions of $X_2$ and $Y_2$
 with respect to excited coherent states will be analytically obtained
\begin{equation}\label{deltax2y2m}
(\Delta X_2)_m^2=(\Delta Y_2)_m^2=\frac{1}{4}s^{4\epsilon}e^{8m\alpha \epsilon^2}(e^{8\alpha \epsilon^2}-1).
\end{equation}
Also, one can calculate the commutator of $X_2$ and $Y_2$ as follows
\begin{equation}\label{x2y2m}
\langle [X_2,Y_2] \rangle_m =\,_\epsilon \langle s,\gamma,m |[X_2,Y_2]
|s,\gamma,m \rangle_\epsilon=\frac{i}{2}s^{4\epsilon}e^{8m\alpha \epsilon^2}(e^{8\alpha \epsilon^2}-1).
\end{equation}
Therefore
\begin{equation}\label{nosqueeze}
(\Delta X_2)_m^2=(\Delta Y_2)_m^2=\frac{1}{2}|\langle [X_2,Y_2] \rangle_m|,
\end{equation}
and obviously no squeezing occurs in amplitude-squared operators, too. Setting $m=0$ in (\ref{deltax2y2m})
and (\ref{x2y2m}) recover (\ref{deltax2y2}) and (\ref{x2y2}), respectively.
    \section{Summary and conclusion}
 In summary, after presenting a brief review on the "coherent state" for
 "quantum systems with continuous spectra" recently introduced by Ben Geloun and Klauder in \cite{geloun}, and at the same time giving a few additional remarkable points, we have introduced the "excited coherent states" for such systems.
 The four axioms of Gazeau-Klauder coherent states and among them the resolution of the identity
 requirement, needed for the over-completeness of the states, are established regarding the introduced exited states. Then, we studied the nonclassicality features of "coherent" and "excited coherent states" by discussing the quantum statistical properties and nonclassicality features of the two mentioned classes of states by evaluating Mandel parameter, second-order correlation function, first and second order squeezing. Interestingly, all of these quantities have been deduced analytically.
 Through this, it is found that both of the coherent and excited coherent states obey sub-Poissonian statistics for
 some values of their variables. We observed that, unlike the photon added coherent states of discrete spectra \cite{agarwal},
 increasing the order of excitation of excited coherent states for continuous spectra reduces the regions and depths of nonclassicality of the states. Even though the two classes of states exhibit neither quadrature squeezing nor amplitude-squared squeezing for all values of their variables, we also found that they can be considered as "intelligent states", in the context of quantum states associated with continuous spectra. \\
 \vspace{0.5cm}\\
 {\bf Acknowledgments}\\
The authors would like to express their utmost thanks to Professor J R Klauder from the University of Florida for valuable comments which helped us to improve the content of the paper. They are also grateful to the referees for their useful comments and suggestions. 

 
 \newpage
 
\begin{figure}
	\centering
		\includegraphics[width=0.80\textwidth]{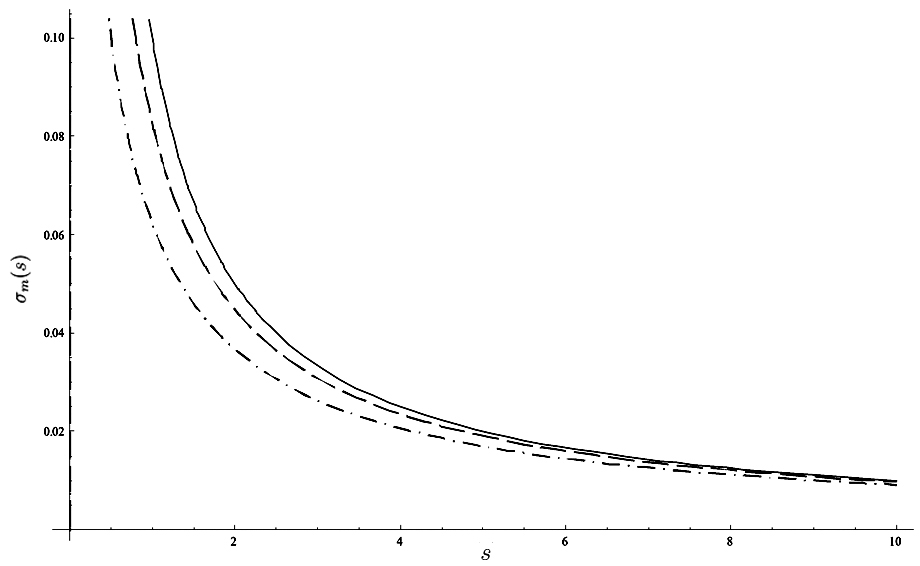}
	\caption{The plots of $\sigma_m(s)$ of excited coherent states for continuous spectra against
  $s$ with $\epsilon=0.07$ and $\alpha=10$ for $m=1$ (dot-dashed curve), $m=3$ (dashed curve) and $m=12$ (solid curve).}
	\label{fig:Figure1}
\end{figure}
\begin{figure}
	\centering
		\includegraphics[width=0.80\textwidth]{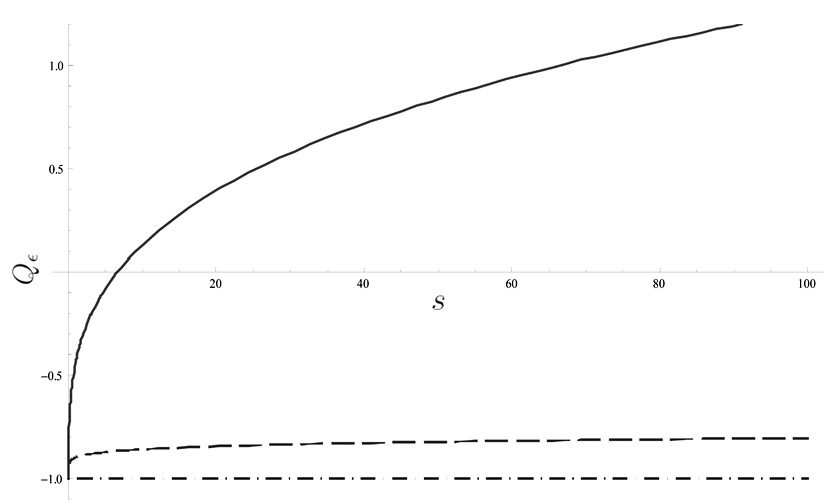}
	\caption{The plots of Mandel parameter of coherent states for continuous spectra against
   $s$ with $\alpha=10$ for $\epsilon=0.01$ (dot-dashed curve), $\epsilon=0.07$ (dashed curve) and $\epsilon=0.15$ (solid curve).}
	\label{fig:Figure2}
\end{figure}
\begin{figure}
	\centering
		\includegraphics[width=0.80\textwidth]{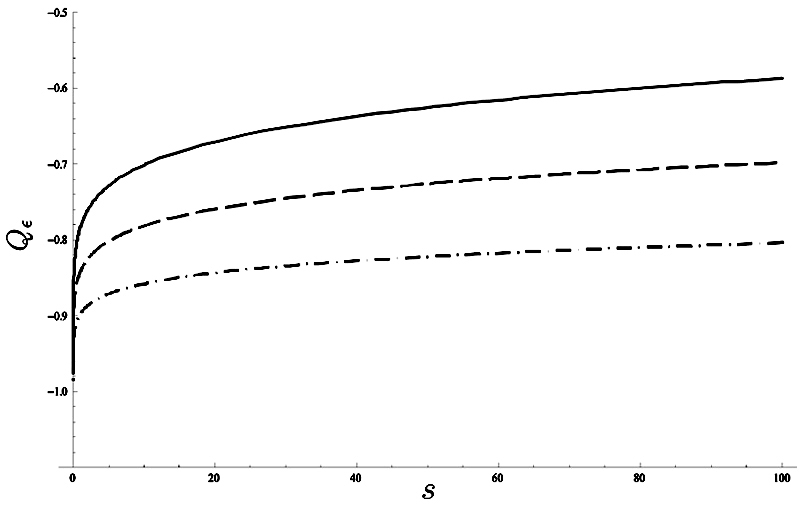}
	\caption{The plots of Mandel parameter of coherent states for continuous spectra against
   $s$ with $\epsilon=0.07$ for $\alpha=10$ (dot-dashed curve), $\alpha=15$ (dashed curve) and $\alpha=20$ (solid curve).}
	\label{fig:Figure3}
\end{figure}
\begin{figure}
	\centering
		\includegraphics[width=0.80\textwidth]{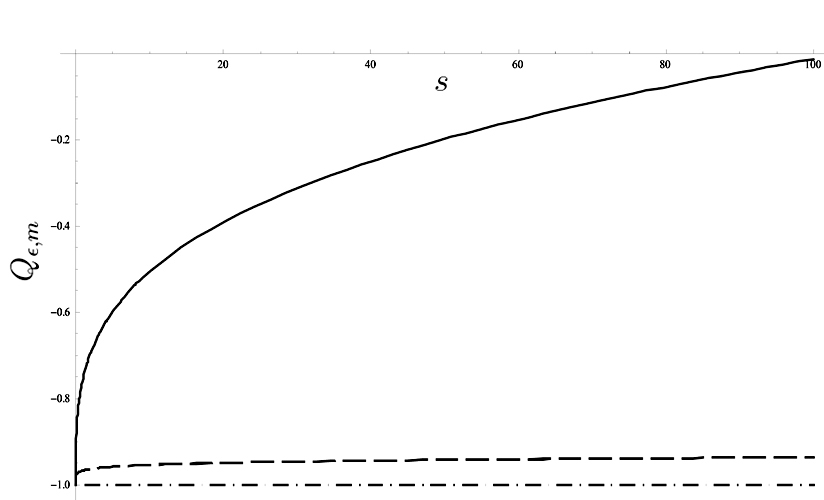}
	\caption{The plots of Mandel parameter of excited coherent states for continuous spectra
  against $s$ with $\alpha=3$ and $m=2$ for $\epsilon=0.01$ (dot-dashed curve),
  $\epsilon=0.07$ (dashed curve) and $\epsilon=0.15$ (solid curve).}
	\label{fig:Figure4}
\end{figure}
\begin{figure}
	\centering
		\includegraphics[width=0.80\textwidth]{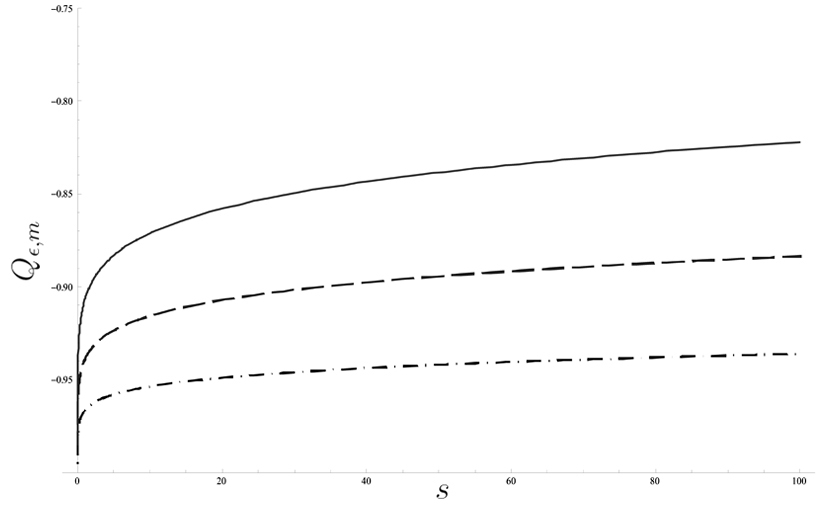}
	\caption{The plots of Mandel parameter of excited coherent states for continuous spectra against $s$ with $\epsilon=0.07$ and $m=2$ for $\alpha=3$ (dot-dashed curve), $\alpha=5$ (dashed curve) and $\alpha=7$ (solid curve).}
	\label{fig:Figure5}
\end{figure}
\begin{figure}
	\centering
		\includegraphics[width=0.80\textwidth]{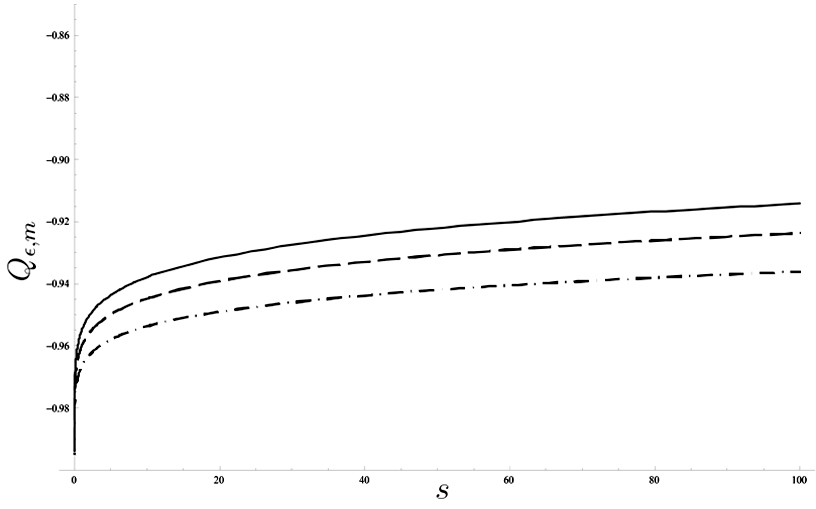}
	\caption{mnbvThe plots of Mandel parameter of excited coherent states for continuous spectra against $s$ with $\epsilon=0.07$ and $\alpha=3$ for $m=2$ (dot-dashed curve), $m=5$ (dashed curve) and $m=7$ (solid curve).}
	\label{fig:Figure6}
\end{figure}

 
 \end{document}